# Material and structural optimization of metal nanowire/AAO composites for high-temperature solar thermal application

Yuefeng Hu, Xiu Li Si, Xiaoling Wu, Guo An Cheng, and Rui Ting Zheng*

*College of Nuclear Science and Technology, Beijing Normal University, Beijing 100875, PR China*

E-mail: huyuefeng@mail.bnu.edu.cn, rtzheng@bnu.edu.cn

## Abstract

In this paper, we study the solar selective absorbing properties of metal nanowire array (NWA) / anodic aluminum oxide (AAO) composites at 1000 K via numerical simulation. The materials and structural parameters which influence the wavelength absorption between 0.28 and 10 microns are simulated and optimized. The results reveal that W NWA / AAO composites with nanowire length of 7.3 microns, fill factor of 0.03, and AAO template thickness of 0.1 microns exhibits the best selective absorption. This composite has 0.90 solar absorptivity and 0.045 thermal emissivity in 1000 K, and shows a photothermal conversion efficiency of 71.5 %.

## 1. Introduction

Solar energy is a kind of environment-friendly, easily-obtained, and cheap energy.[1] Solar-thermal conversion is one of the important ways to utilize the solar energy. Selective absorption coatings (SACs) are an important component in the solar thermal and solar thermo-photovoltaics system, which could capture the most of solar energy in the visible region and minimize radiative heat loss in the long-wavelength range simultaneously.[2,3] However, materials with perfect selective absorption properties do not exist in nature. Combination of different materials, such as metal-dielectric composites, multilayer and semiconductor-metal tandems, are often applied to design such solar selective surfaces.[4-8] In recent years, nano material solar absorber are also investigated. Hu Lu simulated the solar absorption of Si nanowire arrays, found that nanowire arrays can reduce the reflection in a wide wavelength range by changing the filling ratio.[9] Yang obtained extremely low reflectance CNT arrays by adjusting the tube diameter and inter-tube distance.[10] Later carbon nanotube arrays with the spectral selective absorption property were also reported.[11] Veronika reported a high-temperature stable solar absorber based on a metallic 2D photonic crystal, which achieved a thermal transfer efficiency more than 50 % higher than that of a blackbody absorber.[12] Metal nanowire arrays have good thermal conductivities and low emissivity in the long wavelength regime, which makes them potential candidates for solar selective absorber. In previous studies, researchers found that metal nanorod or nanowire array embedded anodic alumina oxide exhibited wonderful spectral selectivity.[13-16] There are lots of reports that achieved high selective absorptivity at low and intermediate temperature.[17-19] However, in practical application, high operating temperature (high than 700 K) is crucial to achieve higher energy conversion

efficiency, especially for solar thermo-photovoltaic applications.[20-22]

In this paper, we simulate the optical properties of five metal NWA / AAO composite and find that tungsten (W) is particularly suitable for high-temperature applications, due to its higher stability at high temperature. Existence of anodic aluminum oxide (AAO) is due to AAO are common used templates for synthesizing metal nanowires. Absorptivity and emissivity of meatal NWAs / AAO composites were simulated by a finite-difference time-domain software (FDTD solution). The influences of microstrcture, such as nanowire length (L), thickness of AAO template (Lt), fill factor of nanowire (FF) were systematically investigated via an optimization software Isight. Optimization results indicate that this nanowire system with 7.3 μm W NWAs, fill factor of 0.03, and thickness of 0.1 μm AAO template exhibits the best solar selective absorption property. With well-established fabrication techniques, this W NWA / AAO composite will have potential applications in high-temperature solar thermal devices or solar thermo-photovoltaics. Our simulation could guide the further design of perfect selective absorbers with periodic subwavelength complex structure.

## 2. Methods and Models

Figure 1 shows the schematic diagram of the NWA / AAO composites. The structure of composites includes metal NWAs, anodic aluminum oxide (AAO) template, and metal basement. The red rectangle (Fig. 1b) exhibits a simulation unit with periodic boundary conditions in this work.

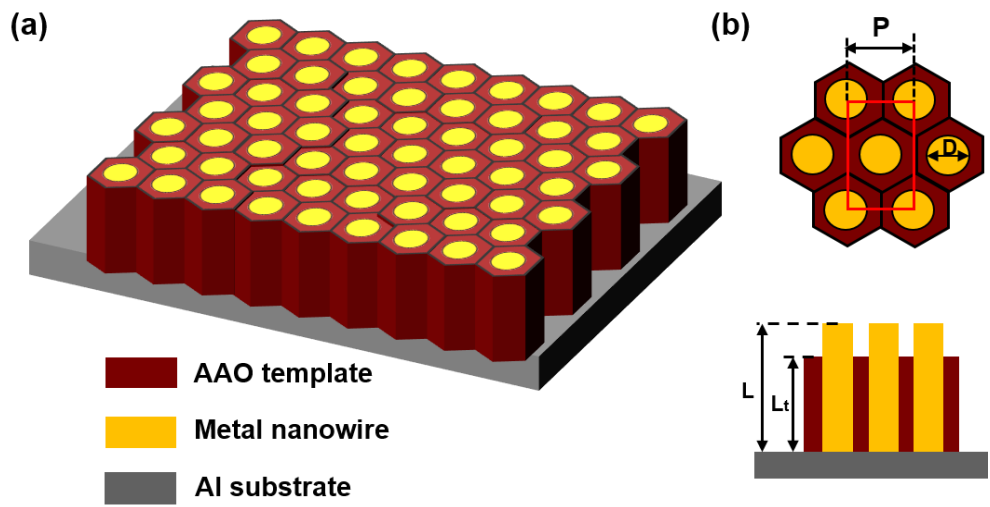

Fig. 1 Schematics of the ordered metal NWA / AAO composites. (a) The composite in a three-dimensional view, these brownish red hexagons mean AAO template, the golden yellow cylinders are metal nanowires and the grey cuboid represents metal substrate. (b) The top view and front view of metal NWA / AAO composites.

The fill factor of metal nanowires is defined as:

$$FF = \frac{\pi}{2\sqrt{3}} \left(\frac{D}{P}\right)^2 \tag{1}$$

Where D is the diameter of nanowires, and P is the pore distance between adjacent nanowires. These golden yellow cylinders in Fig. 1a represent metal NWAs, and the grey plate is the Al substrate. The brownishred honeycomb structure is the AAO template. In this paper, all the optical constants of materials are obtained from *Handbook of Optical Constants of Solids*.[23] The absorption spectrum A(λ) is calculated by A(λ) = 1 - R(λ) - T(λ), where R(λ) and T(λ) are the reflection and transmission spectrum, respectively. Since the metal substrate is semi-infinite, the T(λ) is zero in the whole wavelength region. To characterize the spectral selectivity of metal NWA / AAO composites, A(λ) is used to calculate absorptivity α and emissivity ε according to the following equations:

$$\alpha = \int_\lambda A(\lambda)\Phi(\lambda)d\lambda \bigg/ \int_\lambda \Phi(\lambda)d\lambda \tag{2}$$

$$\varepsilon = \int_\lambda E(\lambda)W(\lambda,T)d\lambda \bigg/ \int_\lambda W(\lambda,T)d\lambda \tag{3}$$

$$W(\lambda,T) = c_1/\lambda^5 \left(e^{c_2/\lambda T} - 1\right) \times 10^{-10} \tag{4}$$

Where Φ(λ) stands for the spectral radiance of AM1.5 (Global tilt).[24] W (λ, T) is the blackbody radiation intensity at the operational temperature (T=1000 K), the range of wavelength varied from 1.5 μm to 10.0 μm is identical with E (λ). E (λ) is the infrared irradiation spectrum. Planck's first and second radiation constant are $c_1$=3.741832e$^{-16}$ Wm$^2$ and $c_2$=1.438768e$^{-2}$ mK. The infrared irradiation of the composite E (λ) equals to the absorption at the thermal equilibrium conditions according to Kirchhoff's laws.[25] The range of λ in the equation (2) goes from 0.28 μm to 1.5 μm, while thermal radiation E (λ) and blackbody radiation W (λ, T))are restricted to a region from 1.5 μm to 10.0 μm. We calculate the photothermal conversion efficiency of composites with equation (5)

$$\eta = \left(\int_\lambda A(\lambda)\Phi(\lambda)d\lambda - \int_\lambda E(\lambda)W(\lambda,T)d\lambda\right) \bigg/ \int_\lambda \Phi(\lambda)d\lambda \tag{5}$$

## 3. Results and Discussions

### 3.1 Influences of materials

Conductive metals have high reflectivity in the visible and infrared regime, because the optical constants of metals are obviously different from that of the air or vacuum. A periodic subwavelength metal structure is proposed to reduce reflection in the short wavelength regime.[26,27] Silver (Ag), copper (Cu), nickel (Ni), tantalum (Ta) and tungsten (W) are suitable metal materials for fabricating selective absorption coatings. Optical properties of six metal NWA / AAO composites are simulated via FDTD solution. Fig. 2 (a) shows the simulated absorption spectra of Ni, Cu, Ag, Ta and W NWA / AAO composites with fixed microstructural parameters (L = 1.0 μm, FF = 0.07, $L_t$ = 1 μm). It could be observed that the absorption spectrum of these metal NWA / AAO composites show a similar trend in wavelength between 280 to 1000 nm. Their absorption in visible and near-infrared region is relatively high, and sharply decreases in the mid and far infrared regime with different cut-off wavelengths. It is well known that the thermal emission enhances rapidly when wavelength exceeds cut-off wavelength at infrared region. For 1000 K, the cut-off wavelength should be no larger than 1.5 μm (based on the blackbody radiation law).

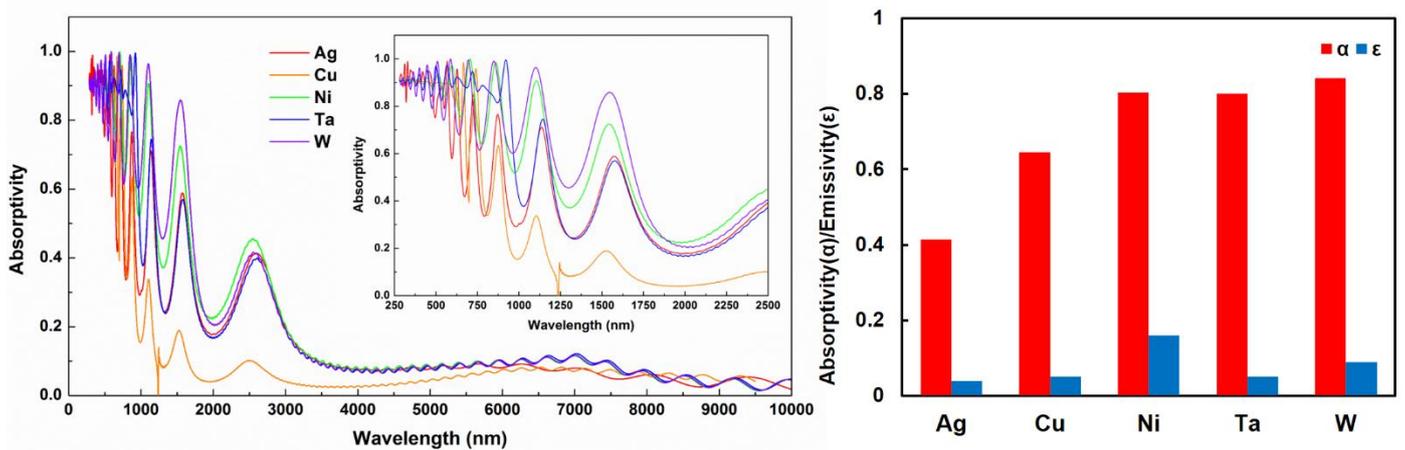

Fig. 2 (a) The absorption of Ag, Cu, Ni, Ta, W NWA / AAO composites with L = 1.0 μm, $L_t$ = 1.0 μm and FF = 0.07. (b) The comparison of α and ε of five ordered metal NWA / AAO composites.

The enhanced absorption of NWA / AAO composites in visible and near-infrared region could be ascribed to two factors, i.e. light scattering effect and surface plasmon polaritons (SPPs). The normal incident light is scattered in a wide angle by the nanowires, increase the optical path of scattering light inside the metal NWA / AAO composites. Meanwhile, SPP, which is the interaction between the surface charges and the light waves illuminated on the surface, enhances the absorption by focusing and concentrating light near to the surface of the metal nanowires.[28] Low absorption in the mid and far infrared region is due to the low absorption coefficient. The difference of absorptivity of different composites within visible and near infrared regime mainly come from the differences of the optical constants between these metals. Fig. 2 (b) shows the calculated absorptivity and emissivity of six metal NWA / AAO composites according equation (2) - (4). The α of metal NWA / AAO composites with the same structural parameters are nearly 0.8 expect Ag and Cu. The reason may be the corresponding extinction coefficients of Ni, W and Ta are larger than that of Cu and Ag. And the lower refractive index of Ag and Cu reduces the scattering effect of NWA/AAO composites. Among those metal NWA / AAO composites, W NWA/AAO composites shows the highest absorptivity (α = 0.842) in visible and near-infrared region and lower emissivity (ε = 0.09) in mid and far infrared region, shows the best spectral selectivity. Furthermore, its superior stability in high temperature makes it be an ideal material of high-temperature selective absorber.

## 3.2 Structure optimization of W NWA / AAO composites

To obtain the best photothermal conversion efficiency, the structure of W NWA / AAO composites needs to be optimized. During our simulation, we chose Latin hypercube sampling method to avoid the interference among structural parameters.[29]

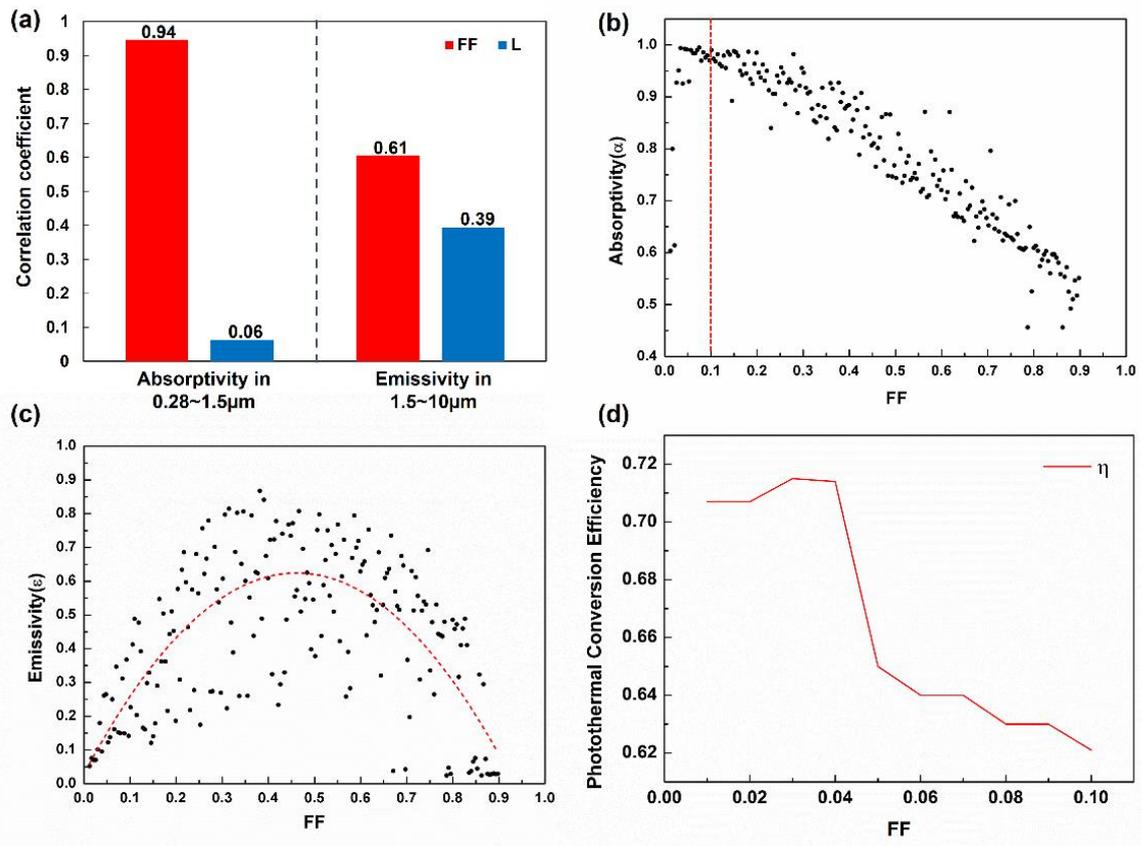

Fig. 3 (a) correlations between parameters and result quantities. (b), (c) α and ε in visible and infrared region of W NWA / AAO composites with FF varying from 0.1 to 0.9. (d) photothermal conversion efficiency of W NWA/AAO composites in 1000 K with FF varying from 0.01 to 0.1.

Generally, emissivity enhancement can be achieved by coupling of resonant electromagnetic modes. The cutoff wavelength is approximately coming from the basic mode of the cylindrical metallic cavity. For radiation with a frequency higher than the fundamental cavity resonant frequency, electromagnetic wave entry into this cavity is forbidden, thus maintaining the desired wavelength selectivity. In this regime, the effective medium theory is valid. The increased surface area between the air and the metal reduces the index contrast, resulting in a slightly lower reflectivity. With these guidelines, we can easily shift the cutoff wavelength by selecting the appropriate D and P, which also can be represented by FF.[30]

Figure 3 (a) shows the correlation between the structural parameters and the absorptivity in W NWA / AAO composites. Among all those parameters, FF plays a key role to the selective absorption of composites. Correlation coefficient between FF and absorptivity reaches 94% in 0.28-1.5 μm region. Moreover, FF also has a 61% correlation coefficient with the emissivity in 1.5-10 μm region. The probable reason for this phenomenon should attribute to the modulation effect of SPP determines by the coupling degree between nanowires (metal) and AAO template (dielectric). FF also represents the contact area of nanowire and AAO template, as mentioned above. When FF is less than 0.1, the absorption shows a trend of reduction with the decrease of FF, shown in the left part on Fig. 3(b). The incident light in visible region with higher energy could escape out of this light trap and scatter into the air when the value of FF is very low, which would result in a blue shift of the cut-off wavelength. This phenomenon will lead to an increasing of infrared irradiation of W NWA / AAO composites. While FF is larger than 0.1, the surface area of W nanowires increases with further increasing of FF, which will result in the stronger reflection and less absorption of solar energy.

Figure 3 (c) shows the ε as a function of FF. When FF is less than 0.45, ε increases with the increasing of FF. The increasing of ε comes from the surface plasmon polariton (SPP). This SPP comes from the resonance interaction between the electromagnetic field of incident light and the surface charge oscillations at the interface of NWA / AAO.[31] The intervals between the nanowires decline when the fill factor enhances, the overlap of SPP resonance field narrows the energy gap, and lead to the enhancement of absorption in the infrared region. When FF is larger than 0.45, the emissivity shows an obvious decreasing tendency. For a sample with large relative surface area of W, the emission properties of composites in infrared region are determined by the optical properties of W, so that the effective refractive index would have a significant enhancement, which could cause a strong reflection in the whole spectrum region. From Fig. 3(d), we could find out that the highest photothermal conversion efficiency can be obtained when the FF equals to 0.03. Taking into account these factors mentioned above, for such selective absorber, W NWA/AAO composites with FF of 0.03 are superior to the others.

The relationship between L, FF and α, ε is shown in Fig.4 (a). The sample designs with better selectivity (high α and low ε) are concentrated in the upper left corner of the graph. So that we could easily find the best design among all these designs. From Fig. 3 (a) and Fig. 4 (a), we can find that the length of nanowires plays an important role in the infrared emissivity of W NWA / AAO composites. Although the effect of AAO (Lt) in the whole system is not shown in the correlation coefficient graph, it is obvious that Lt affects the contact area between W nano arrays and AAO, thus affecting the transmission of SPP. Therefore, the effect of Lt should be discussed simultaneously when analyzing the influence of L. We let L and Lt varying form 0.01μm to 25μm, randomly to get the α and ε for each design, and the impact of these two parameters on α and ε can be observed on Fig. 4. In order to investigate the influence of L and Lt accurately, we design three models shown in Fig.4 (b). We define delta as the difference between L and Lt:

$$delta = L - L_t \qquad (6)$$

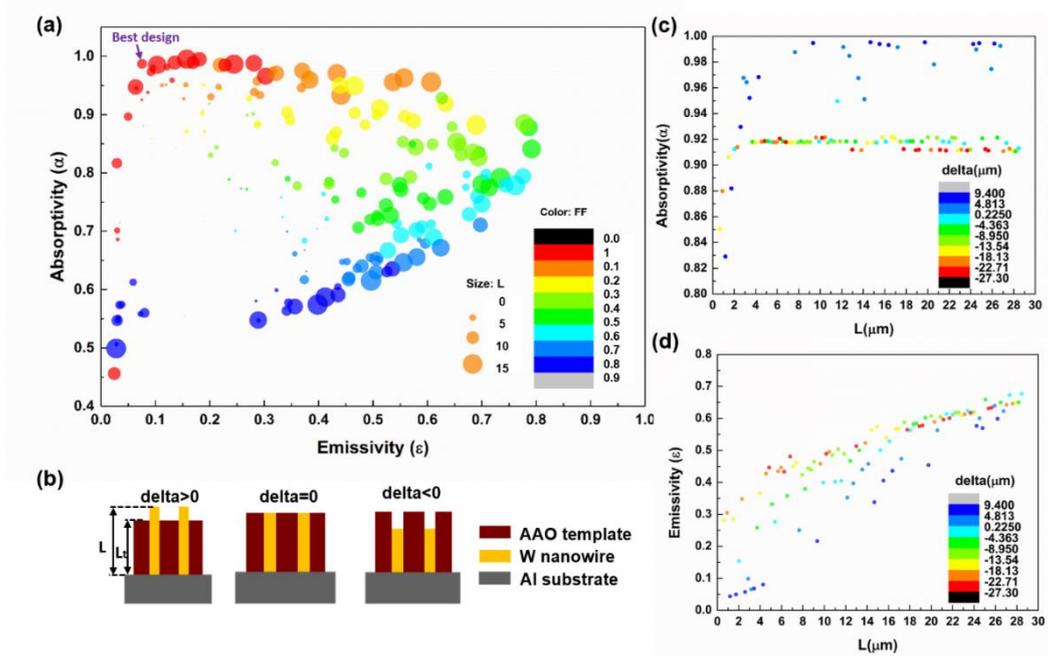

Fig. 4 (a) the relationship between absorptivity, emissivity and the variation of L, FF. (b) front view of simulation model. (c), (d) the changing of α and ε with the variation of L and delta.

The left model is described as L > Lt, the middle one is represented as L = Lt, and the right model shows the circumstance of L < Lt. The result is shown in Fig.4 (c) and (d). When delta ≤ 0, no matter how long the nanowires are, the absorptivity of composite is less than 0.92. The near complete absorption in visible region only exists when delta > 0 (Fig.4(c)), like the left model. When we consider the emissivity in the infrared region, shown in Fig.4 (d), delta > 0 present a lower emissivity than delta ≤ 0. And ε is proportional to L, shorter W nanowire will reduce the emissivity.

The reason why Lt < L cause maximum absorption in visible region is partial comes from presence of interface between metal and air in the composite system. The increasing interface area of air and metal decrease the index contrast of the W NWA / AAO composite, resulting in a higher absorptivity [30]. When the solar spectrum is incident on the surface of the membrane, the charge density of the migratable free electrons at the metal / air and the metal / AAO interface interacts with the incident electromagnetic wave, which makes the charge density of the material fluctuate and result in the collective oscillation of the charge. The surface plasmon is propagated along the interface, enhancing the absorption of the incident light. At the same time, the existence of the variety of interfaces also increases the reflection of the incident light in the material, extending its optical path and increasing the spectral absorption ratio. When Lt > L, the metal / AAO template in the composite material play a role of antireflective-layer. Due to alumina on the short-band of visible region showed a translucent state, which can effectively reduce the direct reflection of the incident light. However, the aluminum oxide film has a certain absorption in infrared region, which will affect the thermal emissivity. It can be seen from the Fig.4 (d) that the thermal emissivity of the composites increases gradually with the increase of the AAO template thickness.

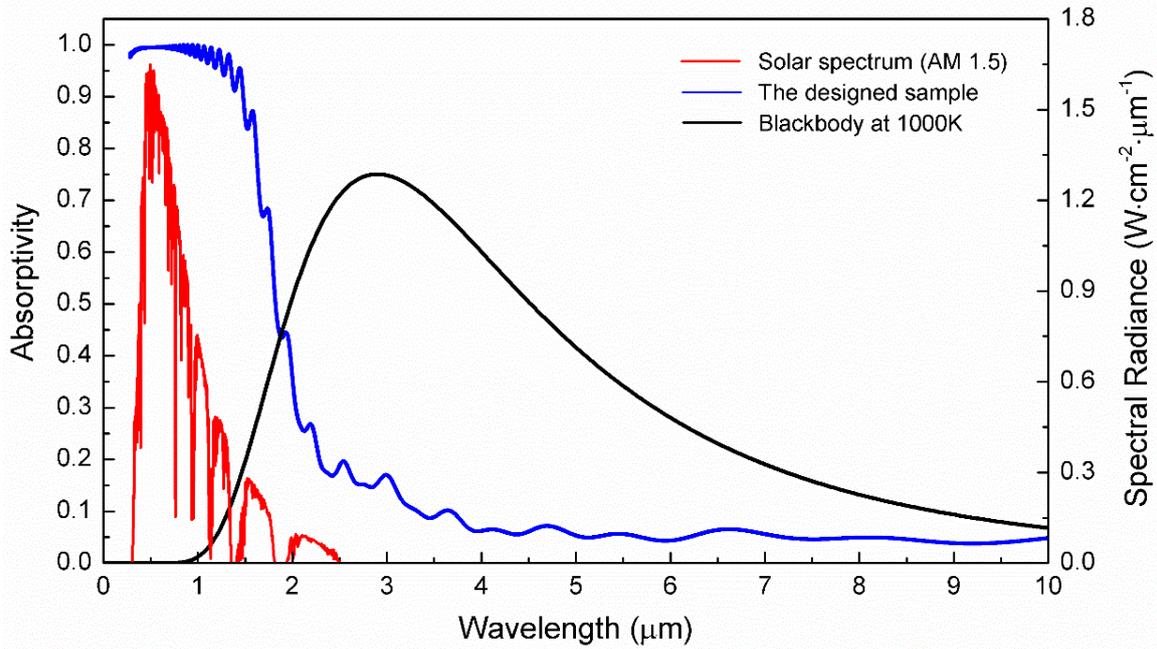

Fig. 5 the simulated spectral of the best designed sample, which has been optimizied for an operating temperature at 1000 K in normal incidence light (blue line), and incident solar spectrum (red line) as well as the emission spectrum of a blackbody at 1000 K (black line).

The normal absorptivity spectrum of the W NWA / AAO composites after optimization is shown in Figure 5. The optimized parameters of this selective absorber are a diameter D = 16 nm, distance between adjacent nanowires P = 174 nm, length of W nanowire L = 7.3 μm, and thickness of AAO Lt = 100nm. As can been seen, the absorptivity in the visible and mid infrared region is high, while shows a steep drop at 2 μm wavelengths. This W NWA / AAO shows an excellent spectrum selectivity at high temperature, and can be used in photothermal conversion.

## Conclusions

We have demonstrated a selective solar absorber based on W NWA / AAO composites for high operation temperature (1000 K) by numerical modeling and optimization. This selective absorber exhibits high absorptivity in solar spectrum and low emissivity in long wavelengths at 1000 K. W and AAO shows good mechanical and optical robustness at high temperature, which makes this composite will have a potential application in solar thermal and thermophotovoltaic devices.

## Acknowledgement

This work is supported by the National Natural Science Foundation of China (11575025) and the Fundamental Research Funds for the Central Universities.